\documentclass[%
 reprint,
 amsmath,amssymb,
 aps,
 prl,
]{revtex4-2}

\usepackage{graphicx}
\usepackage{dcolumn}
\usepackage{bm}
\usepackage{amsmath}
\DeclareMathOperator{\Tr}{tr}

\usepackage{mathtools}
\DeclarePairedDelimiter{\diagfences}{(}{)}
\newcommand{\diag}{\operatorname{diag}\diagfences}

\renewcommand{\Re}{\operatorname{Re}}
\renewcommand{\Im}{\operatorname{Im}}
\renewcommand{\vec}[1]{\boldsymbol{#1}}

\begin{document}


\title{Supershear surface waves reveal prestress and anisotropy of soft materials}

\author{Guo-Yang Li}
\email{gli26@mgh.harvard.edu}
\affiliation{Harvard Medical School and Wellman Center for Photomedicine, Massachusetts General Hospital, Boston, MA 02139, USA}%

\author{Xu Feng}
\affiliation{Harvard Medical School and Wellman Center for Photomedicine, Massachusetts General Hospital, Boston, MA 02139, USA}%

\author{Antoine Ramier}
\affiliation{Harvard Medical School and Wellman Center for Photomedicine, Massachusetts General Hospital, Boston, MA 02139, USA}%

\author{Seok-Hyun Yun}
\email{syun@hms.harvard.edu}
\affiliation{Harvard Medical School and Wellman Center for Photomedicine, Massachusetts General Hospital, Boston, MA 02139, USA}%
 
\date{\today}

\begin{abstract}
Surface waves play important roles in many fundamental and applied areas from seismic detection to material characterizations. Supershear surface waves with propagation speeds greater than bulk shear waves have recently been reported, but their properties are not well understood. In this Letter, we describe theoretical and experimental results on supershear surface waves in rubbery materials. We find that supershear surface waves can be supported in viscoelastic materials with no restriction on the shear quality factor. Interestingly, the effect of prestress on the speed of the supershear surface wave is opposite to that of the Rayleigh surface wave. Furthermore, anisotropy of material affects the supershear wave much more strongly than the Rayleigh surface wave. We offer heuristic interpretation as well as theoretical verification of our experimental observations. Our work points to the potential applications of supershear waves for characterizing the bulk mechanical properties of soft solid from the free surface.
\end{abstract}

\maketitle

Surface wave motion in solids is a classical problem in mechanics, acoustics and seismology, and has found broad applications to nondestructive testing (NDT) of materials \cite{NDT_JAP2006,Garnier2013,WALKER201210}, surface acoustic wave devices \cite{SAWReview_RMP2011,SAWReview_NatMethod2018,SAWSignal_NC2019} and seismic activity monitoring \cite{Seismic_Nature2020,SeismicReview_2018,MisidentificattionSemics_2014}.
The Rayleigh surface wave is the most well studied wave propagating on the free surface of solid. Besides the Rayleigh wave, numerical simulations \cite{HyperelasticityCrack_Nature2003,Buehler2006} and laboratory experiments \cite{CracksInRubber_PRL2004,Mai2020} confirmed the existence of another type of surface wave that propagates faster than bulk shear waves \cite{SupershearRayleighWaves_PRL2013}. This so-called supershear Rayleigh wave allows the surface elastic energy to propagate away with a greater speed and is responsible for supershear crack propagation and supershear earthquake \cite{Earthquake_2007,TheNeedtoStudySpeed_Science2007,SupershearRuptures_Science2013}. Recently, growing needs for mechanical characterization of soft matters, such as elastomers, hydrogels and biological tissues, have spurred renewed interests in surface wave motion \cite{Li2018,Nam2016,Rudykh2014,SuperShearWave_2019APL,SurfaceWaveElastography_JASA2007,OCE_OE2019,Dong2020,Norris2012}. While the Rayleigh surface wave in soft matters is well established and explored for applications, much less is known about supershear waves in soft materials \cite{SuperShearWave_2019APL}. Supershear surface waves are inherently leaky in order to manifest a higher speed than bulk shear waves. Interestingly, supershear surface waves were thought to be supported only in highly viscoelastic materials with high wave attenuation, for example, when the shear wave quality factor (i.e., storage modulus/loss modulus) is less than $6.29$ \cite{ShearWaveWQualityFactor_1992,SupershearRayleighWaves_PRL2013}.

\begin{figure*}[ht!]
    \centering
    \includegraphics{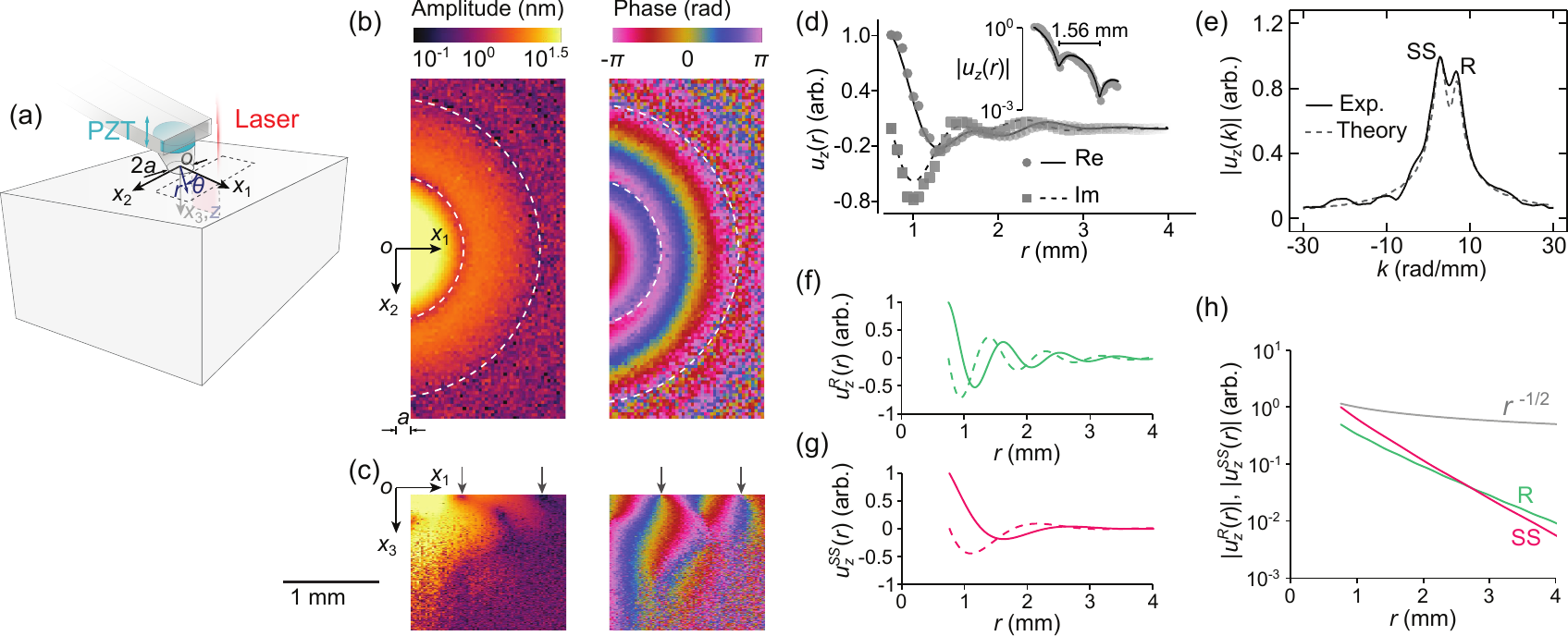}
    \caption{(a) Schematic of the experimental setup. A flat, circular tip connected to a PZT actuator excites mechanical waves at the surface. The wave motion is detected by OCT using a scanning laser beam. (b) the amplitude and phase maps of the top surface. (c) the amplitude and phase maps of a cross-section. The arrows indicate nodes of destructive interference between the supershear and Rayleigh waves.(d) A comparison of the experiment ($f_s = 12$ kHz) and the theoretical model for the surface displacement. Markers, experiment; lines, theory. Inset: the total wave amplitude, which shows the two interference nodes. (e) A comparison of the experiment (solid line) and the theoretical model (dashed lines) in the wavenumber domain. (f) Surface displacements of the Rayleigh wave, $u^R_z(r)$, calculated from the theoretical solution. (g) Surface displacement of the supershear wave $u^{SS}_z(r)$ from the theoretical solution. Solid and dashed lines denote the real and imagery parts, respectively. (h) Experimentally measured amplitudes of the Rayleigh (R) and supershear surface wave (SS) as a function of the radial distance from the tip.}
    \label{Fig1}
\end{figure*}

In this letter, we reveal several underappreciated properties of supershear waves in soft materials via experimental and theoretical investigations. 
Firstly, we show the supershear surface waves can be supported over a broad frequency range, with no restriction on dissipation of the material. 
This invalidates the previous notion that supershear surface waves are only supported in highly dissipative materials. 
We visualize supershear surface waves in three dimensions in transparent rubber using optical coherence elastography. 
The measured wave profiles are in excellent agreement with our theoretical model. 
Secondly, we show that the stress on the soft material alters the speed of the supershear surface wave opposite to what is commonly known for Rayleigh surface waves. 
We account for this apparently surprising result using the acoustoelastic theory, finite-element (FE) simulation, and simple heuristic explanation. 
Extending this finding, we show how material anisotropy affects supershear surface waves distinctly different from Rayleigh surface waves. 

In Fig. 1(a) we show the schematic of our experiment. The sample used in this study is a silicone rubber (EcoFlexTM 00-50, Smooth-On Inc.). The approximate size is 8$\times$4 cm in the lateral extent and 4 cm along depth, which is large enough to avoid wave reflections at the edges. 
Mechanical waves were excited at the surface by a vertically vibrating, flat tip with a circular contact area with a radius $a$ of $\sim 0.75$ mm, which is driven by a PZT actuator (Thorlabs, PA4CEW).
The driving frequency of the PZT, $f_s$, was step-tuned from $2$ to $20$ kHz. The vertical component of the displacement (on the order of $10$ nm) was measured by using a home-built sweep-source optical coherence tomography (OCT) system \cite{OCE_OE2019}. 
The OCT system measures the optical phase delay modulation of a reflected laser beam at different lateral locations ($\sim 8$ ms per location) with a sampling rate of $\sim 43$ kHz. By conducting Fourier transformation of the acquired phase trace, we obtain the amplitude and phase of the displacement (Supplemental Material). 
Figures \ref{Fig1}(b)-(e) show a typical wave field measured when $f_s = 12$ kHz. 
The minimums in wave amplitude and fluctuations in phases (as indicated by the dashed lines and arrows) represent destructive interference between the Rayleigh and supershear surface waves.

\begin{figure}[bt!]
    \centering
    \includegraphics[width=0.48\textwidth]{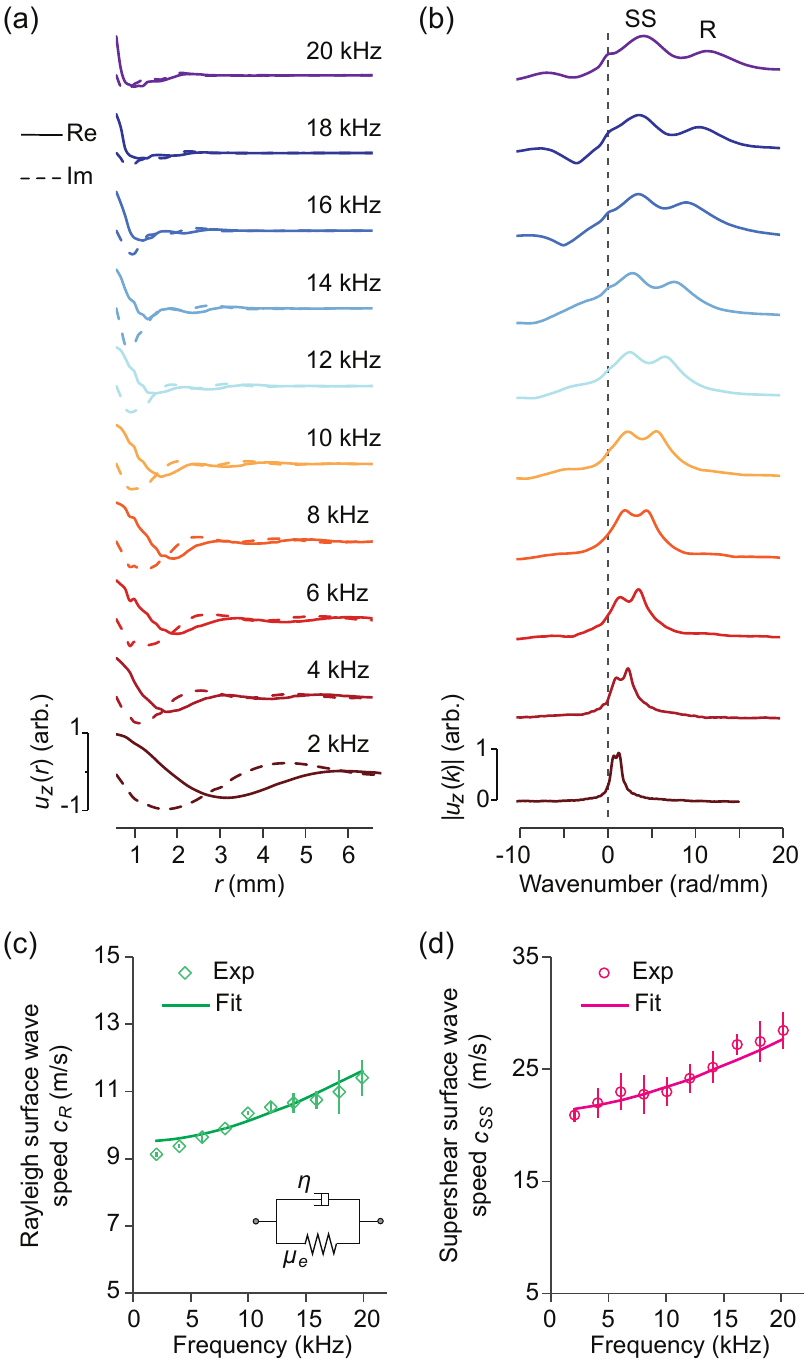}
    \caption{Surface waves measured over a frequency range from $2$ to $20$ kHz. (a) The real and imagery parts of the surface displacement. (b) The Fourier transformation of the surface displacement to resolve the supershear (SS) and Rayleigh (R) surface waves. (c-d) Phase velocities of the Rayleigh and supershear surface waves. Error bar: standard deviations over five measurements. Solid curves, the Kelvin–Voigt model with $\mu_e = 101.3$ kPa and $\eta = 0.65 $ Pa$\cdot$s (inset in (c)).}
    \label{Fig2}
\end{figure}

To understand the observation, we perform a theoretical study on the near-field displacement of the surface waves. Consider a semi-finite solid which occupies the space $z \geq 0$, as shown in Fig. \ref{Fig1}(a). 
Since the problem is axisymmetric, we introduce a cylindrical coordinate system $\left( r,\theta,z \right)$ and suppose the tip imposes a uniform time-harmonic pressure $p(r,t)=p_0e^{i2\pi f_s t}$ ($r \leq a$) to the contact surface, where $p_0$ denotes the amplitude of the pressure, $a$ the radius of the cylindrical tip, and $t$ the time. 
The equilibrium equation for isotropic materials is $(\lambda + 2\mu)\nabla\nabla \cdot \vec{u} - \mu \nabla \times \nabla \times \vec{u} = \rho \partial^2 \vec{u} / \partial t^2$, where $\rho$ is the density. 
$\lambda$ and $\mu$ are Lam\'e constants, and $\lambda \gg \mu$ for soft materials. Assuming a time-harmonic wave $\vec{u} = \vec{u}_0e^{i 2 \pi f_s t}$, we get an explicit equation for $z$ component of the surface displacement $u_{0z}(r>a)$ (Supplementary Material):
\begin{equation}
\begin{split}
    & u_{0z}(r)  =  i \pi \frac{a p_0 k^2_S}{\mu} \left[ \frac{J_1(k_R a) k_R}{\mathcal{F}'(k_R)}H_0^{(1)}(-k_R r) \right.\\
             & \left. + \frac{J_1(k_{SS} a) k_{SS}}{\mathcal{F}' (k_{SS})}H_0^{(1)}(-k_{SS} r) \right] = u^R_z(r) + u^{SS}_z(r),
\end{split}
\label{eq:analytical solution}
\end{equation}
where $R$ and $SS$ in the sub- and super-scripts indicate the Rayleigh and supershear waves, respectively, $J_1$ is the Bessel function of the first kind, and $H_0^{(1)}$ is the Hankel function of the first kind. For large $r$, $ H_0^{(1)}(-k r) \rightarrow e^{-ikr}r^{-1/2}$; Equation (\ref{eq:analytical solution}) describes the two propagating waves with wavenumbers, $k_R$ and $k_{SS}$, which are the solutions of a secular equation:
\begin{equation}
\begin{split}
    \mathcal{F}(k) & = (2k^2 - k_S^2)^2 \\ 
    & - 4k^2\sqrt{k^2(k^2 - k_S^2)} \cdot \mathrm{sign}\{\Re(k^2 - k_S^2)\},
\end{split}
\label{eq:F(k)}
\end{equation}
where $k_S = 2 \pi f_s \sqrt{\rho / \mu}$ is the wavenumber of the bulk shear wave, and $\mathcal{F}' = d\mathcal{F}/dk$. The two physical solutions are $k_R = 1.047k_S$ and 
\begin{equation}
k_{SS} = (0.4696 - 0.1355i) k_S.
\label{eq:k_SS}
\end{equation}
Here the negative sign in the imaginary coefficient results in an exponential decay of the supershear wave amplitude over $r$.

The phase velocity $c$ of a wave is related to its wavenumber $k$, via $c = 2 \pi f_s / \Re(k)$. For pure elastic materials with a real value of $k_S$, the supershear wave is leaky with a complex value of $k_{SS}$ and has $c_{SS} = 2.13 c_{T0}$, where $c_{T0} = [\Re(\sqrt{\rho/\mu})]^{-1}$ is the phase velocity of the bulk shear wave. For general viscoelastic materials, we find that the ratio $c_{SS}/c_{T0}$ increases slightly as the shear-wave quality factor $Q$ ($=\Re(\mu) / \Im(\mu)$) decreases; for example, it is 2.18 at $Q = 6$ and 2.99 at $Q = 0$.

The two terms in Eq. (\ref{eq:analytical solution}) results in the interference patterns observed in our experiments. To quantitatively compare the theoretical model and experiments, in Fig. \ref{Fig1}(d) we show the experimental normalized displacement data (markers) along the radial direction at $f_s = 12$ kHz and also plot the theoretical curves ordained by fitting the experimental data with Eq. (\ref{eq:analytical solution}) with $k_S$ as the only fitting parameter.
The best fitting gives $k_S \approx 6.93 - 1.03i$ mm$^{-1}$. Using $\rho \approx 1,070 $ kg/m$^3$, we obtain the storage modulus $\mu' \approx 110.8$ kPa and the loss modulus $\mu'' \approx 33.7$ kPa ($\mu = \mu' + i\mu''$). 
By performing the Fourier transformation we move the experimental data from the spatial domain to the wavenumber domain (see Fig. \ref{Fig1}(e)) and observe two peaks that correspond to the Rayleigh and supershear surface waves, respectively. 
The theoretical model shows a good agreement with the experiment in the wavenumber domain. Strictly speaking, an inverse Hankel transformation is required to extract the wavenumber from the displacement profiles. However, We find that the Fourier transform is a convenient approximation with errors less than 4$\%$ for all our experimental cases (see Supplementary Material, Fig. S1). 

In Figs. \ref{Fig1}(f) and (g) we plot the experimentally measured surface displacements $u^R_z(r)$ and $u^{SS}_z(r)$. As shown in Fig. \ref{Fig1}(h)), the wave amplitudes decrease exponentially over $r$, much steeper than the $r^{-1/2}$ dependence expected for radially propagating surface waves in pure elastic materials \cite{USinSolid_Roes}. The fast exponential decay is due to the relatively low $Q$ of the rubber sample with $\mu'/\mu''$ = 3.3 at 12 kHz.

We further investigated the effect of viscoelasticity by varying the stimulus frequency. 
Figure \ref{Fig2}(a) and (b) show the near-field surface displacements obtained at $f_s$ from 2 to 20 kHz in the space and wavenumber domains, respectively (see Supplementary Material, Fig. S2, for the comparison with the theoretical solution). The relative amplitudes of the two modes are slightly frequency dependent. This is because the excitation efficiency of the wave is sensitive to the size, $a$, of pressure loading. From Eq. (\ref{eq:analytical solution}), for $a$ = 0.75 mm we find that $|u^{SS}_z(a)| > |u^{R}_z(a)|$ when $f_s > \sim $5 kHz.

\begin{figure*}[hbt!]
    \centering
    \includegraphics{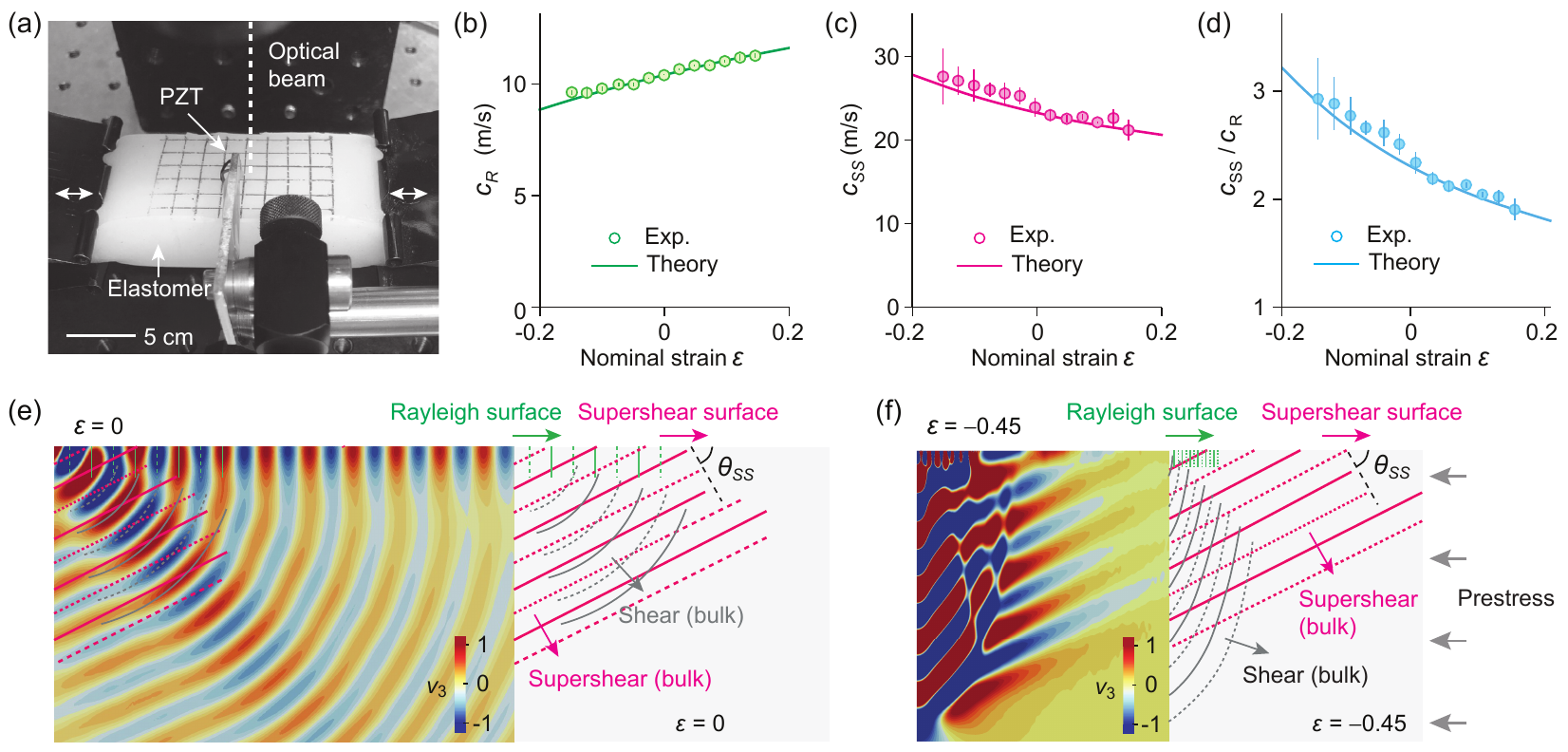}
    \caption{(a) Uniaxial extension setup. (b)-(c) Variations of the phase velocities of the Rayleigh and supershear surface waves, and (d) their ratio at different nominal strain values $\varepsilon$. Circles, experimental data measured at $f_s =$ 12 kHz; error bars, standard deviation of five measurements. Solid lines, theoretical solutions with $\mu = 119$ kPa and $\zeta = 0$. (e)-(f) Finite element simulations of the wave field (left) and illustrations of different waves (right) in a soft material under (e) stress-free and (f) compressive ($\varepsilon = -0.45$) conditions. $\theta_{SS}$ denotes the propagation angle of the supershear surface wave in the medium (cyan).}
    \label{Fig3}
\end{figure*}

From the measured wavenumbers, the phase velocities of the two waves are computed. The results are shown in Fig. \ref{Fig2}(c) and (d). 
The velocities were measured to increase with frequency. This indicates the viscoelasticity of the material. The experimental frequency dependence is fit well into the Kelvin-Voigt model using $\mu = \mu_e + i 2\pi f_s \eta$, where $\mu_e = 101.3$ kPa and $\eta = 0.65$ Pa$\cdot$s (see inset in Fig. \ref{Fig2}(c)).
It has been previously proposed that the supershear surface wave was supported only when $Q < 6.29$ \cite{ShearWaveWQualityFactor_1992}. 
However, at $f_s = 2$ kHz we measure $k_S \approx 1.25 - 0.052i$ mm$^{-1}$ from the experimental data using Eq. (\ref{eq:analytical solution}) and find $Q \approx 12.1$. Certainly, Eq. (\ref{eq:analytical solution}) predicts the existence of the supershear surface wave without restriction on the quality factor in soft materials \cite{SupershearRayleighWaves_PRL2013,ComplexConjugateRoots_JASA2001}.

Typical soft materials including rubber has nonlinear elastic properties or hyperelasticity. Both Young's modulus along the direction of strain and shear modulus facing the direction of the strain increase with the magnitude of strain. In these hyperelastic materials, an extensional prestress increases the velocity of waves propagating along the direction of stretch, and compression decreases the wave speed. This general characteristic has been observed and well documented for the Rayleigth surface wave both experimentally and theoretically. However, little is known about the effect of prestress on the supershear surface wave. 

To describe the supershear wave, we use the acoustoelastic theory in continuum mechanics \cite{Acoustoelasticity,AcoustoelasticRayleigh_1981} and consider a nominal uniaxial stress $\vec N_{11}$ in a lateral coordinate, $x_1$, and its resulting nominal strain $\varepsilon$ along the direction. For isotropic, incompressible, hyperelastic materials, the deformation gradient tensor is given by $\vec F = \diag{1+\varepsilon,(1+\varepsilon)^{-1/2},(1+\varepsilon)^{-1/2}}$. Among several different hyperelastic models, we consider the Mooney-Rivlin model that is widely used for rubbery materials under modest deformation ($0 < \varepsilon < 100\%$) \cite{KIM_MRmodel_2011,Fitting_hyperelasticmodel}. It describes the strain energy function as
    $W(\vec F) = (\mu/2) \left [\zeta \left( I_1 - 3 \right) + (1 - \zeta)\left( I_2 - 3 \right) \right].$
Here, $\mu$ is shear modulus of the material at the stress-free condition, $\zeta$ is a material-dependent parameter ($0 \leq \zeta \leq 1$), $I_1 = \Tr{(\vec{C})}$, and $I_2 = \left[ I_1^2 -  \Tr{(\vec{C}^2)}\right]/2$, where $\vec{C} = \Tr{(\vec{F}^{\mathrm{T}} \vec{F})}$ is the right Cauchy-Green deformation tensor. The nominal stress is related to the energy function by $\vec N = \partial W / \partial \vec F - \Bar{p} \vec F^{-1}$, where $\Bar{p}$ is the Lagrange multiplier for the incompressibility constraint (or pressure to maintain constant volume). For 
a small uniaxial stress, $N_{11} \approx 3\mu \varepsilon [1 - (2 - \zeta)\varepsilon]$.

The constitutive equation for acoustoelastic waves in the pre-stressed, deformed material can be written with the Eulerian elasticity tensor $\vec{\mathcal A^0}$ defined as $\mathcal A^0_{piqj} = F_{p \alpha} F_{q \beta} \partial^2 W/ \partial F_{i \alpha} \partial F_{j \beta}$, where the subscripts $i,j,p,q, \alpha, \beta$ represent different coordinates. The infinitesimal elastic wave $\vec u$ in the deformed material coordinates is governed by the incremental dynamic equation \cite{Ogden2007,SurfaceWavePrestressed1990}: 
    $\mathcal A^0_{piqj}\partial^2 u_j/ \partial x_p \partial x_q - \partial \hat{\Bar{p}} / \partial x_i = \rho \partial^2 u_i / \partial t^2$,
where the Einstein summation convention has been adopted. 
$\hat{\Bar{p}}$ denotes the increment of $\Bar{p}$. Utilizing the plane wave solution of the dynamic equation and the stress free boundary conditions at the surface ($x_3 = 0$), we obtain the following secular equation:
\begin{equation}
\begin{split}
    \gamma (\alpha -\gamma & -\rho \mathcal{C}^2)  + (2 \beta + 2 \gamma - \rho \mathcal{C}^2) \times  \\
    & \left[ \gamma (\alpha - \rho \mathcal{C}^2) \right]^{\frac{1}{2}} \mathrm{sign}\{\Re(\alpha - \rho \mathcal{C}^2)\} = 0,
\end{split}
\label{eq:secular equation for surface waves with prestress}
\end{equation}
where $\alpha = \mu [\zeta (1+\varepsilon)^2 + (1-\zeta)(1+\varepsilon)]$, $\gamma = \mu[\zeta/(1+\varepsilon) + (1-\zeta)/(1+\varepsilon)^2]$, $\beta = (\alpha+\gamma)/2$, and $\mathcal{C} = 2 \pi f_s / k$. 
In the absence of prestress ($\vec N=0$), $\alpha = \beta = \gamma = \mu$, and Eq. (\ref{eq:secular equation for surface waves with prestress}) reduces to $\mathcal{F}(k) = 0$ in Eq. (\ref{eq:F(k)}) (see Supplementary Material). The solutions of the equation $\mathcal{C}$ yield the phase velocities of the surface waves via $c = \left[\Re(\mathcal{C}^{-1}) \right]^{-1}$. We obtain the following result for small $\varepsilon$:
\begin{equation}
\begin{split}
    c_R & \approx 0.955\, c_{T0} (1+ 0.65 \varepsilon + 0.49\zeta\varepsilon -0.37 \varepsilon^2  + 0.21\zeta \varepsilon^2), \quad \\
    c_{SS} & \approx 2.13 \, c_{T0} (1 - 0.77 \varepsilon +0.51\zeta\varepsilon + 1.05 \varepsilon^2 - 0.56\zeta\varepsilon^2),
\end{split}
\label{eq:first_order_approximation}
\end{equation}
where $c_{T0} = \sqrt{\mu/\rho}$ is the shear wave speed in the stress-free configuration. 
Note that the coefficient of the linear $\varepsilon$ term for the supershear wave is always negative for $\zeta = [0,1]$.

Using a custom-built mechanical setup and the OCT system (Fig. \ref{Fig3}(a) and Supplementary Material), we measured the velocities of the Rayleigh and supershear surface waves in a rubber sample at different magnitudes of prestress. Figure \ref{Fig3} (b)-(d) shows the experimental results for a range of $\varepsilon$ from $-$0.2 to 0.2. The Rayleigh wave velocity increases with lateral strain as expected. Interestingly, we find that the supershear wave velocity actually decreases with the strain, the opposite behavior. The best fit to Eq. (\ref{eq:first_order_approximation}) was obtained with $\zeta \approx 0$ (Supplemental Material), which is shown in Fig. \ref{Fig3}. No specific physical interpretation of $\zeta = 0$ in the phenomenological Mooney-Rivlin model is known. The ratio of $c_{SS}/c_R \approx 2.23 - 3.30 \varepsilon + 5.46 \varepsilon^2$ is uniquely related to strain.

To understand the negative sensitivity of $c_{SS}$ to uniaxial extension, we performed FE simulation (Supplemental Material). Figure \ref{Fig3}(e)
illustrates wave motion in the $x_1$-$x_3$ cross-sectional plane in the stress-free configuration. Three distinct waves from the excitation point are seen: the Rayleigh wave with phase planes normal to the surface, the supershear wave with phase planes tilt at a angle $\theta_{SS}$, and a spherical shear wave. The supershear wave is a leaky surface wave, whose energy is radiated into the medium in the form of a planar shear wave. This shear wave has a speed of $c_{T0}$. The leaky angle $\theta_{SS}$ satisfies Snell's law: $c_{SS} \cos(\theta_{SS}) = c_{T0}$, from which $\theta_{SS}=62^\circ$. The supershear surface wave can be viewed as a bulk shear wave created at the surface and propagating with the steep angle into the medium. Now, we have a qualitative explanation for the negative dependence of $c_{SS}$ on $\varepsilon$. As the medium is stretched in $x_1$, it is compressed in $x_3$ by the Poisson's ratio. This compression decreases wave speeds along $x_3$, just like compression in $x_1$ decreases wave speeds along $x_1$. Since the propagation direction of the supershear wave is more vertical than horizontal, the effect of prestress on $c_{SS}$ is opposite to that of $c_R$. Figure \ref{Fig3}(f) shows the result of FE simulation at a prestress condition ($\varepsilon = -0.45$). The compressive stress decreases the shear wave speed along $\theta = 0^{\circ}$ direction but increases the shear wave speed along $\theta \approx 62^\circ$.
Under uniaxial prestress, the plane shear wave speed becomes anisotropic with angle-dependent velocity \cite{Auld_1973}:
\begin{equation}
c_T = \sqrt{(\alpha \cos^4\theta + 2\beta \sin^2 \theta \cos^2 \theta + \gamma \sin^4\theta)/\rho}, 
\label{eq:C_T}
\end{equation}
and $c_{SS} =$ $c_T(\theta_{SS})$ / $\cos(\theta_{SS})$. Over $-0.5 \le \varepsilon \le 0.5$, we find the variation of $\theta_{SS}$ is $< 1.4^{\circ}$  (Supplementary Material). Therefore, $c_{SS} \approx 2.13 \,c_T(\theta = 62^\circ, \varepsilon)$. Whereas the speed of the Rayleigh surface wave $c_R \approx$ 0.955 $c_T(\theta = 0^\circ, \varepsilon)$.

\begin{figure}[hbt!]
    \centering
    \includegraphics[width=0.45\textwidth]{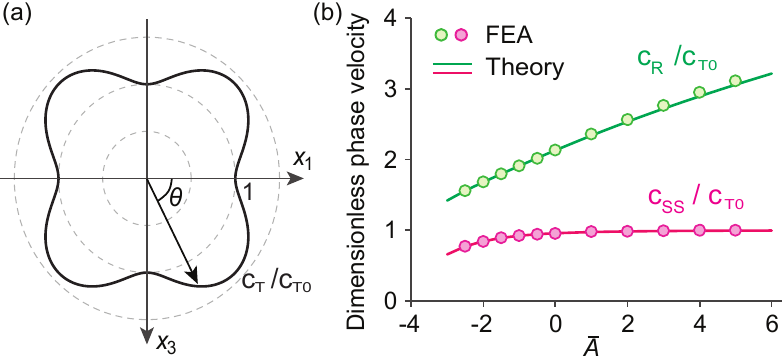}
    \caption{(a) Normalized phase velocity in an anisotropic material ($\Bar{A} = 4$). (b) The dependence of surface wave speeds on material anisotropy.}
    \label{Fig4}
\end{figure}

From the angular interpretation, we expect $c_{SS}$ to be sensitive to material anisotropy. In an orthotropic material with the symmetry orientation aligned with the coordinate system, the shear wave speed is given by Eq. (\ref{eq:C_T}) with $\alpha=\gamma=C_{55}$ and $2\beta =$ $C_{11} + C_{33} - 2C_{13} - 2C_{55}$, where $C_{ij}$ are components of the stiffness matrix \cite{Auld_1973} (Supplemental Material). We introduce an anisotropy index $\Bar{A} = E^*_1/\mu_{13} - 4$, where $E^*_1 = (C_{11} + C_{33} - 2C_{13})$ is the plane strain Young’s modulus, and $\mu_{13} = C_{55}$, a shear modulus. Then, $c_T(\theta) =$ $c_{T0} [\, (\cos^4\theta$ + $(\Bar{A} +2 )$ $\sin^2 \theta \cos^2 \theta$ + $\sin^4\theta)\,]^{1/2}$, where $c_{T0} = \sqrt{\mu_{13}/\rho}$ (see Fig. \ref{Fig4}(a)). For isotropic materials, $\Bar{A} = 0$ and $c_T(\theta) = c_{T0}$. 
As shown in Fig. \ref{Fig4}, the ratio of the supershear surface wave velocity to the bulk shear speed shows a strong dependence on $\Bar{A}$ while the ratio of the Rayleigh surface wave speed to the shear wave speed remains nearly unchanged, especially when $\Bar{A} > 0$. The leaky angle $\theta_{SS}$ has a weak dependence on $\Bar{A}$ (Supplementary Material). Our FE simulations agree well with the theoretical curves (Fig. \ref{Fig4}(b)).

In conclusion, we have reported on the properties of the supershear surface wave in soft materials. We provided theoretical and experimental descriptions of the effects of viscoelasticity, prestress and anisotropy of the material on the velocity of the supershear wave.
The specific properties of the supershear surface wave are quite distinct from those of the Rayleigh surface waves. The difference gives us a new opportunity to use the two wave modes to characterize the material anisotropy or mechanical stress \cite{Li_JASA2020} of bulk materials from measurements at their free surface.
One possible application is to characterize soft biological tissues using the two surface waves with OCT elastography \cite{OCE_SR2020}. 
The destructive interference of the supershear and Rayleigh surface waves in the near field may be applied to trap and manipulate particle across length scales \cite{Single_Beam_NC2020}.
Finally, our findings may be useful in the investigations of supershear dynamics of solids, such as ultra-fast dynamic ruptures, where the stress concentration gives rise to the hyperelastic stiffening.

This study was supported by grants from National Institutes of Health (P41-EB015903, R01-EY027653, DP1-EB024242).

%

\end{document}



\title{Supporting Information \\ for \\ \textit{Supershear surface waves reveal prestress and anisotropy of soft materials}}

\date{\today}

\author{Guo-Yang Li}
\email{gli26@mgh.harvard.edu}
\affiliation{Harvard Medical School and Wellman Center for Photomedicine, Massachusetts General Hospital, Boston, MA 02139, USA}%

\author{Xu Feng}
\affiliation{Harvard Medical School and Wellman Center for Photomedicine, Massachusetts General Hospital, Boston, MA 02139, USA}%

\author{Antoine Ramier}
\affiliation{Harvard Medical School and Wellman Center for Photomedicine, Massachusetts General Hospital, Boston, MA 02139, USA}%

\author{Seok Hyun Yun}
\email{syun@hms.harvard.edu}
\affiliation{Harvard Medical School and Wellman Center for Photomedicine, Massachusetts General Hospital, Boston, MA 02139, USA}%

\maketitle

\renewcommand{\thepage}{S\arabic{page}} 
\renewcommand{\thesection}{S\arabic{section}}  
\renewcommand{\thetable}{S\arabic{table}}  
\renewcommand{\thefigure}{S\arabic{figure}}
\renewcommand{\theequation}{S-\arabic{equation}} 

\clearpage
\section{Mechanical model for surface wave excitation with a time harmonic stimulus}\label{Sec.S1}
As shown in Fig. 1(a), let the plane $z = 0$ be the boundary of the semi-infinite soft solid which occupies the space $z \geq 0$. A harmonic pressure $p(r,t)=p_0e^{i \omega t}$ is applied on a circular region ($r \leq a$) of the free surface where $p_0$ is the amplitude of the pressure and $t$ denotes the time. $\omega = 2 \pi f_s$, where $f_s$ is the stimulus frequency. Since the problem is axisymmetric we have $\vec{u} = \vec{u}_0e^{i \omega t}$, where $\vec{u}_0 = u_{0r} \vec{e}_r + u_{0z} \vec{e}_z$. $\mathbf{e}_r$ and $\mathbf{e}_z$ denote the unit vectors along $r$ and $\theta$ directions, respectively. Inserting $\vec{u}$ into the equilibrium equation $(\lambda + 2\mu)\nabla\nabla \cdot \vec{u} - \mu \nabla \times \nabla \times \vec{u} = \rho \partial^2 \vec{u} / \partial t^2$, we can get
\begin{equation}
    (\lambda + 2\mu)\nabla\nabla \cdot \vec{u}_0 - \mu \nabla \times \nabla \times \vec{u}_0 = -\rho \omega^2 \vec{u}_0,
\label{eq:equilibrium equation}    
\end{equation}
where $\rho$ is the density. $\lambda$ and $\mu$ are Lam\'e constants and $\lambda \gg \mu$ for soft solids studied here.

To solve this problem, we introduce $\phi=\nabla \cdot \vec{u}_0$ and $\psi \vec{e}_{\theta} = \nabla \times \vec{u}_0$, where $\vec{e}_{\theta}$ denotes the unit vector along $\theta$ direction.
\begin{equation}
    \phi = \frac{1}{r}\frac{\partial (r u_{0r})}{\partial r} + \frac{\partial u_{0z}}{\partial z}, \quad \psi = \frac{\partial u_{0r}}{\partial z} - \frac{\partial u_{0z}}{\partial r}. 
\label{eq:potential and stream functions}    
\end{equation}
Inserting Eq. (\ref{eq:potential and stream functions}) into Eq. (\ref{eq:equilibrium equation}) we can obtain
\begin{equation}
    (\lambda + 2\mu) \frac{\partial \phi}{\partial z} - \frac{\mu}{r}\frac{\partial (r\psi)}{\partial r} + \rho \omega^2 u_{0z} = 0, \quad (\lambda + 2\mu)\frac{\partial \phi}{\partial r} + \mu \frac{\partial \psi}{\partial z} + \rho \omega^2 u_{0r} = 0. 
\label{eq:equilibrium equation-2}    
\end{equation}
Eliminating $\phi$ and $\psi$ from Eq. (\ref{eq:equilibrium equation-2}) respectively yields following decoupled equilibrium equations  
\begin{equation}
    \frac{1}{r}\frac{\partial}{\partial r}\left( r\frac{\partial \phi}{\partial r} \right) + \frac{\partial^2 \phi}{\partial z^2} + k_L^2\phi = 0, \quad \frac{\partial}{\partial r}\left[ \frac{1}{r} \frac{\partial}{\partial r}(r\psi) \right] + \frac{\partial^2 \psi}{\partial z^2} + k_S^2 \psi = 0,
\label{eq:equilibrium equation-3}    
\end{equation}
where $k_L = \omega / \sqrt{(\lambda + 2\mu)/\rho}$ and $k_S = \omega / \sqrt{\mu/\rho}$.

The in-plane components of the Cauchy stress $\vec{\sigma}$, expressed in terms of $\phi$ and $\psi$, are (neglecting the time harmonic terms)
\begin{equation}
    \frac{\rho \omega^2}{\mu^2}\sigma_{zz} = \frac{2}{r}\frac{\partial}{\partial r}\left( r\frac{\partial \psi}{\partial z} \right) - \frac{\nu^2(\nu^2 - 2)}{r}\frac{\partial}{\partial r}\left( r \frac{\partial \phi}{\partial r} \right) - \nu^4\frac{\partial^2 \phi}{\partial z^2}, \quad
    \frac{\rho \omega^2}{\mu^2}\sigma_{zr} = \frac{\partial}{\partial r}\left[ \frac{1}{r} \frac{\partial(r\psi)}{\partial r} \right] - \frac{\partial^2 \psi}{\partial z^2} - 2\nu^2\frac{\partial^2 \phi}{\partial r \partial z},    
\label{eq:Cauchy stress}    
\end{equation}
where $\nu = k_S/k_L$. The boundary conditions on the free surface are
\begin{equation}
\begin{split}
    \sigma_{zz} = p_0 (r \leq a), \quad \sigma_{zz} = 0 (r > a), \quad \sigma_{zr} = 0,    
\end{split}
\label{eq:BCs}    
\end{equation}
where we again neglect the time harmonic terms. Inserting Eq. (\ref{eq:BCs}) into Eq. (\ref{eq:Cauchy stress}) we thus get the boundary conditions expressed in terms of $\phi$ and $\psi$.

To solve Eq. (\ref{eq:equilibrium equation-3}), here we follow the Miller's work \cite{Miller_1954} and perform Hankel transformations (from $r$ to $k$)  
\begin{equation}
    \frac{\mathrm{d}^2 \Bar{\phi}_0}{\mathrm{d} z^2} - (k^2 - k_L^2)\Bar{\phi}_0 = 0, \quad \frac{\mathrm{d}^2 \Bar{\psi}_1}{\mathrm{d} z^2} - (k^2 - k_S^2)\Bar{\psi}_1 = 0,
\label{eq:equilibrium equation-4}    
\end{equation}
where ${\Bar{\phi}}_0$ and ${\Bar{\psi}}_1$ are the 0 and 1-order Hankel transformations of $\phi$ and $\psi$, i.e.,
\begin{equation}
    \Bar{\phi}_0(k,z) = \int_{0}^{\infty} \phi(r,z) r J_0(k r) \mathrm{d}r, \quad \Bar{\psi}_1(k,z) = \int_{0}^{\infty} \psi(r,z) r J_1(k r) \mathrm{d}r,
\label{eq:Hankel transformation}    
\end{equation}
where $J_0$ and $J_1$ are the Bessel function of the first kind of order 0 and 1, respectively. Similarly, we perform Hankel transformations of Eqs.(\ref{eq:Cauchy stress}) and (\ref{eq:BCs}) to get
\begin{equation}
    \frac{\rho \omega^2}{\mu^2}\Bar{\sigma}_{zz,0}  = - \nu^4\frac{\partial^2 \Bar{\phi}_0}{\partial z^2} + 2k \frac{\mathrm{d}\Bar{\psi}_1}{\mathrm{d}z} + \nu^2(\nu^2 - 2) k^2 \Bar{\phi}_0, \quad
    \frac{\rho \omega^2}{\mu^2}\Bar{\sigma}_{zr,1}  = -\left( \frac{\mathrm{d}^2 \Bar{\psi}_1}{\mathrm{d}z^2} - 2\nu^2 k \frac{\mathrm{d}\Bar{\phi}_0}{\mathrm{d}z} + k^2 \Bar{\psi}_1 \right),    
\label{eq:Cauchy stress-2}    
\end{equation}
and 
\begin{equation}
    \Bar{\sigma}_{zz,0}  = \frac{a p _0 J_1(k a)}{k}, \quad
    \Bar{\sigma}_{zr,1}  = 0,    
\label{eq:BCs-2}    
\end{equation}
where $\Bar{\sigma}_{zz,0}$ and $\Bar{\sigma}_{zr,1}$ are the 0 and 1-order Hankel transformations of $\sigma_{zz}$ and $\sigma_{zr}$, respectively.

Solving Eq.(\ref{eq:equilibrium equation-4}) with the boundary conditions given by Eqs. (\ref{eq:Cauchy stress-2}) and (\ref{eq:BCs-2}), for Rayleigh surface wave we get
\begin{equation}
    \Bar{\phi}_0 = \frac{\rho \omega^2 (k_S^2 - 2 k^2)}{\nu^2 \mu^2 k \mathcal{F}(k)} a p_0 J_1(k a)e^{-z\sqrt{k^2 - k_L^2}},\quad 
    \Bar{\psi}_1 = \frac{2 \rho \omega^2 \sqrt{k^2 - k_L^2}}{\mu \mathcal{F}(k)} a J_1(k a)e^{-z\sqrt{k^2 - k_S^2}},
\label{eq:solution-potential and stream}    
\end{equation}
where
\begin{equation}
    \mathcal{F}(k) = (2k^2 - k_S^2)^2 - 4k^2\sqrt{k^2 - k_L^2}\sqrt{k^2 - k_S^2}.
\label{eq:secular equation}    
\end{equation}
In Eq. (\ref{eq:solution-potential and stream}), we have taken $e^{-z\sqrt{k^2-k^2_S}}$ for $\Bar{\psi}_1$ to make sure $\Bar{\psi}_1 \rightarrow 0$ when $z \rightarrow + \infty$. For supershear surface wave, we should take $e^{z\sqrt{k^2-k^2_S}}$ for $\Bar{\psi}_1$ \cite{Xu2013}, which results in a plus sign for the second term of $\mathcal{F}(k)$. To get a universal form for the secular equation, we replace the equation (\ref{eq:secular equation}) with
\begin{equation}
    \mathcal{F}(k) = (2k^2 - k_S^2)^2 - 4k^2\sqrt{k^2 - k_L^2}\sqrt{k^2 - k_S^2} \cdot \mathrm{sign}\{\Re(k^2 - k_S^2)\}.
\label{eq:secular equation-modified}
\end{equation}

Substitution of Eq. (\ref{eq:solution-potential and stream}) in Hankel transformation of Eq. (\ref{eq:potential and stream functions}), we can get the Hankel transformations of $u_{0z}$ and $u_{0r}$
\begin{equation}
\begin{split}
    \Bar{u}_{z,0} & = \frac{a p_0 J_1(k a) \sqrt{k^2 - k_L^2}}{\mu k \mathcal{F}(k)} \left[ 2k^2e^{-z\sqrt{k^2 - k_S^2}} + (k_S^2 - 2k^2)e^{-z\sqrt{k^2 - k_L^2}} \right],\\ 
    \Bar{u}_{r,1} & = \frac{a p_0 J_1(k a)}{\mu \mathcal{F}(k)} \left[ 2\sqrt{k^2 - k_L^2}\sqrt{k^2 - k_S^2}e^{-z\sqrt{k^2 - k_S^2}} + (k_S^2 - 2k^2)e^{-z\sqrt{k^2 - k_L^2}} \right].    
\end{split}
\label{eq:Hankel tranformation of displacment}    
\end{equation}
The displacement $u_{0z}$ and $u_{0r}$ can be obtained by performing inverse Hankel transformations to Eq. (\ref{eq:Hankel tranformation of displacment}).

Here we focus on $u_z$, which is measured in our experiments. 
\begin{equation}
    u_z(r,z) = \frac{a p_0}{\mu}\int_{0}^{\infty} \frac{J_1(k a) \sqrt{k^2 - k_L^2}}{\mathcal{F}(k)} \left[ 2k^2e^{-z\sqrt{k^2 - k_S^2}} + (k_S^2 - 2k^2)e^{-z\sqrt{k^2 - k_L^2}} \right]J_0(k r) \mathrm{d}k. 
\label{eq:inverse Hankel tranformation of displacment}    
\end{equation}
Accordong to Royston et al.\cite{Royston_JASA1999}, an equivalent form of Eq.(\ref{eq:inverse Hankel tranformation of displacment}) is
\begin{equation}
    u_z(r,z) = \frac{a p_0 i}{\pi \mu}\int_{-\infty}^{\infty} \frac{J_1(k a) \sqrt{k^2 - k_L^2}}{\mathcal{F}(k)} \left[ 2k^2e^{-z\sqrt{k^2 - k_S^2}} + (k_S^2 - 2k^2)e^{-z\sqrt{k^2 - k_L^2}} \right]K_0(-i k r) \mathrm{d}k,
\label{eq:inverse Hankel tranformation of displacment-2}    
\end{equation}
where $K_0$ is the modified Bessel function of the second kind. The superiority of Eq. (\ref{eq:inverse Hankel tranformation of displacment-2}) is that the Cauchy principal value theorem is applicable. There are some comprehensive discussions on how to deal with the contributions from the poles and the branch cuts \cite{graff2012wave,Achenbach_JASA2002,ComplexConjugateRoots_JASA2001} when performing the integration in Eq. (\ref{eq:inverse Hankel tranformation of displacment-2}). Since we are only interested in the surface waves, we can neglect the contributions of the branch integrals and only consider the contributions of the residues \cite{graff2012wave}. Therefore we can get
\begin{equation}
    u_z(r,0) = \frac{2 a p_0 k_S^2}{\mu} \sum_{k} \frac{J_1(ka)\sqrt{k^2 - k_L^2}}{\mathcal{F}'(k)}K_0(ikr),
\label{eq:inverse Hankel tranformation of displacment-3}    
\end{equation}
where $\mathcal{F}' = \partial \mathcal{F} / \partial k$, and $k$ denotes the root of $\mathcal{F}(k) = 0$. 

For soft materials ($\lambda \gg \mu$) we considered in this study, $k, k_S \gg k_L$. In this case Eq. (\ref{eq:secular equation-modified}) reduce to 
\begin{equation}
    \mathcal{F}(k) = (2k^2 - k_S^2)^2 - 4k^2\sqrt{k^2(k^2 - k_S^2)}\cdot \mathrm{sign}\{\Re(k^2 - k_S^2)\},
\label{eq:secular equation-2}    
\end{equation}
Taking the two roots of Eq.(\ref{eq:secular equation-2}) that correspond to the Rayleigh and supershear surface waves, i.e., $k_R = 1.047k_2$ and $k_{SS} = (0.4696-0.1355i)k_2$, we get the Eq. (1) in the main text, where we have used the Hankel function of the first kind $H_0^{(1)}(x)$ to take the place of $K_0(x)$ (note that $K_0(x) = i\pi/2 H_0^{(1)}(ix)$). The other root of of Eq.(\ref{eq:secular equation-2}) has a positive imagery part, which denotes an exponential increase in wave amplitude along the radial direction. This root thus is excluded.

\begin{figure}[bt!]
    \centering
    \includegraphics[width=.3\textwidth]{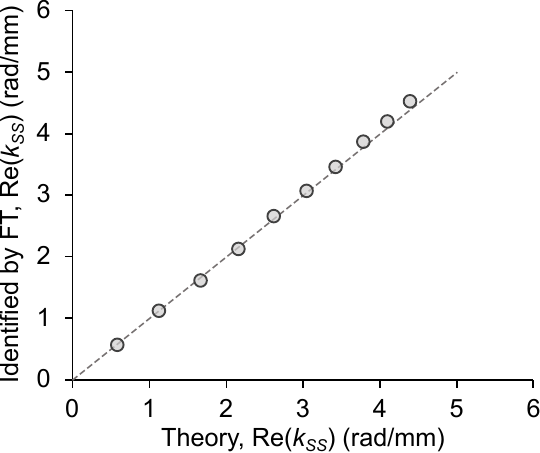}
    \caption{Comparison between $\Re(k_{SS})$ obtained using Fourier transformation (FT) and the theory value. The relative error is less than 4\%.}
    \label{FigS0}
\end{figure}

\begin{figure}[bt!]
    \centering
    \includegraphics[width=0.9\textwidth]{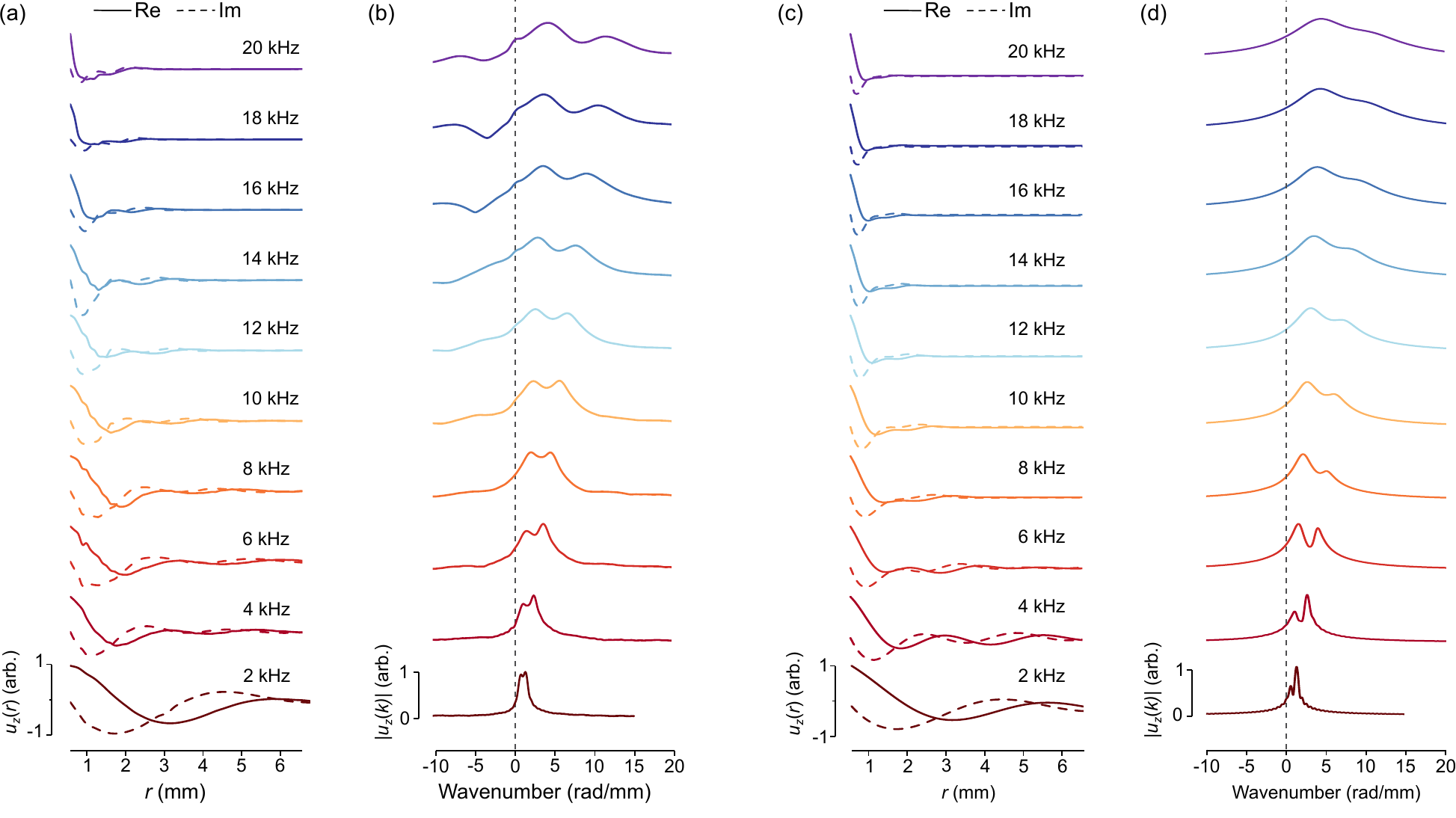}
    \caption{Comparison between the experiment and theory. (a)-(b) Experimental data. (c)-(d) Theoretical results obtained with the Kelvin–Voigt model, $\mu_e = 101.3$ kPa and $\eta = 0.65 $ Pa$\cdot$s.}
    \label{FigS1}
\end{figure}

\clearpage
\section{Acoustoelastic model}\label{Sec.S2}
Here the incremental dynamic theory is briefly revisited and then adopted to derive the secular equation for the surface waves that incorporates the effect of prestress. Consider, a nominal stress $\vec N$ homogeneously deforms the sample from the stress-free configuration to the current configuration in which the surface waves propagate. The infinitesimal elastic wave $\vec u$ in the current configuration is governed by the incremental dynamic equation \cite{Ogden2007} 
\begin{equation}
    \mathcal A^0_{piqj}\partial^2 u_j/ \partial x_p \partial x_q - \partial \hat{\Bar{p}} / \partial x_i = \rho \partial^2 u_i / \partial t^2,
\label{eq:incremental dynamics}
\end{equation}
where the Einstein summation convention has been adopted. $\vec{\mathcal A^0}$ is the Eulerian elasticity tensor and defined as $\mathcal A^0_{piqj} = F_{p \alpha} F_{q \beta} \partial^2 W/ \partial F_{i \alpha} \partial F_{j \beta}$ ($i,j,p,q, \alpha, \beta = 1,2,3$), where $W(\vec F)$ is the strain energy function, $F_{i \alpha} = \partial x_i / \partial X_\alpha$ is the deformation gradient, and $x_i$ and $X_\alpha$ denote the Cartesian coordinates of the points in the current and stress-free configurations. The nominal stress is related to the strain energy function by $\vec N = \partial W / \partial \vec F - \Bar{p} \vec F^{-1}$, where $\Bar{p}$ is the Lagrange multiplier for the incompressibility constraint. $\hat{\Bar{p}}$ denotes the increment of $\Bar{p}$. Denote the principle stretch ratios by $\lambda_i$ ($i = 1,2,3$), then we have $\vec F = \diag{\lambda_1,\lambda_2,\lambda_3}$. For the incompressible materials we are interested in this study, $\lambda_1 \lambda_2 \lambda_3 = 1$ and
\begin{equation}
    \nabla \cdot \vec{u} = 0.
    \label{eq:incompressibility}
\end{equation}
 
We consider the plane wave in $x_1-x_3$ (i.e., $u_2 = 0$ and $\partial ()/ \partial x_2 = 0$). From Eq. (\ref{eq:incompressibility}) we can introduce a stream function $\chi(x_1,x_3,t)$ such that: $u_1 = \partial \chi / \partial x_3$ and $u_3 = -\partial \chi / \partial x_1$. Inserting these into Eq. (\ref{eq:incremental dynamics}) and eliminating $\hat{\Bar{p}}$ we can get
\begin{equation}
    \alpha \chi_{,1111} + 2\beta \chi_{,1133} + \gamma \chi_{,3333} = \rho (\chi_{,11tt} + \chi_{,33tt}),
    \label{eq:incremental dynamics-2D}
\end{equation}
where $()_{,i}$ and $()_{,t}$ denote the partial deviates with respect to coordinate $x_i$ ($i=1,3$) and $t$, and
\begin{equation}
    \alpha = \mathcal A^0_{1313},\quad 2\beta = \mathcal A^0_{1111} + \mathcal A^0_{3333} - 2\mathcal A^0_{1133} - 2 \mathcal A^0_{3113},\quad \gamma = \mathcal A^0_{3131}.
    \label{eq:incremental parameters}
\end{equation}

\subsection{Surface wave}
On the surface ($x_3 = 0$) the stress-free boundary conditions apply. So the incremental nominal stress, denoted by $\hat{\vec N}$, which can be expressed with $\chi$ by \cite{Ogden2007}
\begin{equation}
    \hat{N}_{31} = - \gamma \chi_{,11} + \gamma \chi_{,33}, \quad  \hat{N}_{33,1} = \rho \chi_{,3tt} - (2\beta + \gamma)\chi_{,113} - \gamma \chi_{,333},
    \label{eq:incremental BCs}
\end{equation}
should be zeros. 

To derive the secular equation for surface wave, we should take 
\begin{equation}
    \chi = \Bar{\chi}e^{-skx_3}e^{i(\omega t - kx_1)},
    \label{eq:stream function for surface wave}
\end{equation}
where $s$ is a dimensionless attenuation and $\Bar{\chi}$ is a constant amplitude. Substitution of Eq.(\ref{eq:stream function for surface wave}) into Eq.(\ref{eq:incremental dynamics-2D}) yields
\begin{equation}
    \gamma s^4 - (2\beta - \rho \mathcal{C}^2)s^2 + \alpha - \rho \mathcal{C}^2 = 0,
    \label{eq:dimensionless attenuation}
\end{equation}
where $\mathcal{C} = \omega / k$. 

Here we use $s_1$ and $s_2$ to denote the two roots of Eq. (\ref{eq:dimensionless attenuation}) that have positive real parts. Similar as the discussion in Sec. S1, for Rayleigh surface wave we take
\begin{equation}
    \chi = \left( \Bar{\chi}_1e^{-s_1 k x_3} + \Bar{\chi}_2e^{-s_2 k x_3} \right)e^{ik(\mathcal{C} t - x_1)},
    \label{eq:stream function for surface wave-2}
\end{equation}
where
\begin{equation}
   s_1^2 + s_2^2 = (2\beta - \rho \mathcal{C}^2) / \gamma,\quad s_1^2s_2^2 = (\alpha - \rho \mathcal{C}^2) / \gamma
   \label{eq:dimensionless attenuation-2}
\end{equation}
according to Eq. (\ref{eq:dimensionless attenuation}).
Substitution of Eq. (\ref{eq:stream function for surface wave-2}) into Eq.(\ref{eq:incremental BCs}) yields the following linear equations of $\Bar{\chi}_1$ and $\Bar{\chi}_2$
\begin{equation}
    (s_1^2 + 1)\Bar{\chi}_1 + (s_2^2 + 1)\Bar{\chi}_2 = 0, \quad \left[2\beta + \gamma - \rho \mathcal{C}^2 - \gamma s_1^2 \right] s_1 \Bar{\chi}_1 + \left[2\beta + \gamma - \rho \mathcal{C}^2 - \gamma s_2^2 \right] s_2 \Bar{\chi}_1 = 0.
    \label{eq:linear equations of the attenuation}
\end{equation}
To have nontrivial solutions for Eq. (\ref{eq:linear equations of the attenuation}) we must have
\begin{equation}
    \gamma(\alpha - \gamma - \rho \mathcal{C}^2) + (2\beta + 2\gamma - \rho \mathcal{C}^2)\left[ \gamma(\alpha - \rho \mathcal{C}^2)\right]^{\frac{1}{2}} = 0.
    \label{eq:acoustoelastic secular equation}
\end{equation}
In the derivation of Eq. (\ref{eq:acoustoelastic secular equation}) we have used Eq. (\ref{eq:dimensionless attenuation-2}). 

For the supershear surface wave, we should take, without loss of generality, $s_1$ and $-s_2$. In this way we get a sign change in Eq. (\ref{eq:acoustoelastic secular equation}). Similar as Eq. (\ref{eq:secular equation-2}) we get
\begin{equation}
\begin{split}
    \gamma (\alpha -\gamma -\rho \mathcal{C}^2)  + (2 \beta + 2 \gamma - \rho \mathcal{C}^2) \left[ \gamma (\alpha - \rho \mathcal{C}^2) \right]^{\frac{1}{2}} \mathrm{sign}\{\Re(\alpha - \rho \mathcal{C}^2)\} = 0,
    \label{eq:acoustoelastic secular equation-modified}
\end{split}
\end{equation}

Eq. (\ref{eq:acoustoelastic secular equation-modified}) is the secular equation that incorporates the effect of the prestress, from which the phase velocity of the surface wave $c$ can be obtained by $c = \left[\Re(\mathcal{C}^{-1}) \right]^{-1}$.

While different notations has been adopted, Eq. (\ref{eq:acoustoelastic secular equation-modified}) will reduce to Eq. (\ref{eq:secular equation-2}) in the absence of prestress. In the stress-free configuration $\alpha = \beta = \gamma = \mu$, where $\mu$ is the linear shear modulus. Recall the definitions $\mathcal{C} = \omega / k$ and $k_S = \omega / \sqrt{\mu/\rho}$, we can rewrite Eq. (\ref{eq:acoustoelastic secular equation}) as
\begin{equation}
    -k_S^{-2}k^{-2} + \left( 4k_S^{-2} - k^{-2} \right) \left[ k_S^{-2} \left( k_S^{-2} - k^{-2} \right) \right]^{\frac{1}{2}} = 0,
    \label{eq:acoustoelastic secular equation-stress free}
\end{equation}
which is equivalent to $\mathcal{F}(k) = 0$ after some simple derivations.

\subsection{Shear wave}
To obtain the wave speed for bulk shear wave, we can take
\begin{equation}
    \chi = \Bar{\chi}e^{ik(c_T t - x_1\cos{\theta} - x_3\sin{\theta})},
    \label{eq:stream function for shear wave}
\end{equation}
where $\theta$ denotes the angle between the wave propagation direction and $x_1$ axis. Substitution of Eq. (\ref{eq:stream function for shear wave}) into Eq.(\ref{eq:incremental dynamics-2D}) leads to
\begin{equation}
    \rho c_T^2 = \alpha \cos^4{\theta} + 2\beta \cos^2{\theta} \sin^2{\theta} + \gamma \sin^4{\theta}.
    \label{eq:bulk shear wave speed}
\end{equation}

\subsection{Results for Mooney-Rivlin model}
To capture the material response, we introduce the incompressible Mooney-Rivlin model
\begin{equation}
    W_{\mathrm{MR}} = \frac{\mu}{2} \left [\zeta \left( I_1 - 3 \right) + (1 - \zeta)\left( I_2 - 3 \right) \right],
\label{eq:M-R model}
\end{equation}
where $0 \leq \zeta \leq 1$ is a dimensionless parameter. $I_1 = \Tr{(\vec{C})}$ and $I_2 = \left[ I_1^2 -  \Tr{(\vec{C}^2)}\right]/2$, where $\vec{C} = \Tr{(\vec{F}^{\mathrm{T}} \vec{F})}$ is the right Cauchy-Green deformation tensor. The Mooney-Rivlin model reduce to the neo-Hookean model \cite{Fitting_hyperelasticmodel} when $\zeta = 1$. 

With Eq. (\ref{eq:M-R model}) we get
\begin{equation}
    \alpha = \mu \left[ \zeta \lambda_1^2 + (1 - \zeta) \lambda_3^{-2}\right], \qquad \gamma = \mu \left[ \zeta \lambda_3^2 + (1 - \zeta) \lambda_1^{-2}\right], \qquad \beta = (\alpha + \gamma)/2,
\label{eq:incremental_parameters_MR}
\end{equation}
where $\lambda_1 = 1 + \varepsilon$ and $\lambda_2 = \lambda_3 = (1+\varepsilon)^{-1/2}$ for the uniaxial stretch loading. $\varepsilon$ denotes the nominal strain. The corresponding nominal stress is
\begin{equation}
    N_{11} = \mu \left[\zeta + (1 - \zeta)/(1 + \varepsilon) \right] \left[1 + \varepsilon - (1 + \varepsilon)^{-2} \right] = 3\mu \varepsilon \left[ 1 - (2 - \zeta)\varepsilon\right] + O(\varepsilon^3),
\label{eq:nominal stress_MR}
\end{equation}

Inserting Eq. (\ref{eq:incremental_parameters_MR}) into the secular equation (\ref{eq:acoustoelastic secular equation}) we get the wave speeds of the Rayleigh $c_R$ and supershear $c_{SS}$ surface waves.

\begin{figure}[hbt!]
    \centering
    \includegraphics[width=.3\textwidth]{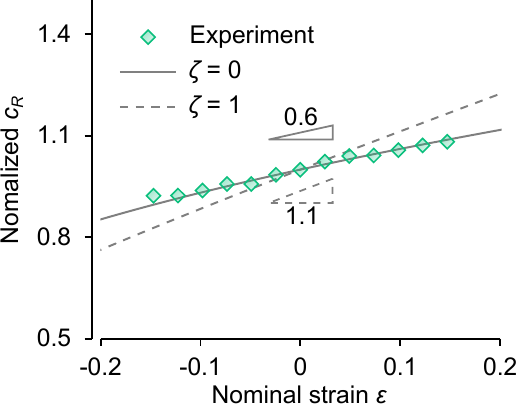}
    \caption{Variation of the normalized wave speed of Rayleigh surface wave ($c_R$) when changing the nominal strain $\varepsilon$.}
    \label{FigS2}
\end{figure}

The first-order approximation of $c_R$ is
\begin{equation}
    c_R = (0.955 + 0.615\varepsilon)\sqrt{\mu/\rho} \quad \mathrm{for} \quad \zeta = 0, \qquad c_R = (0.955 + 1.09\varepsilon)\sqrt{\mu/\rho} \quad \mathrm{for} \quad \zeta = 1.
\label{eq:Rayleigh surface wave speed-first order}
\end{equation}
As shown in Fig. \ref{FigS2}, the experimental data approximately follows the theoretical curve given by $\zeta = 0$. We thus used $\zeta = 0$ in Fig. 4 of the main text. 

\begin{figure}[hbt!]
    \centering
    \includegraphics[width=.3\textwidth]{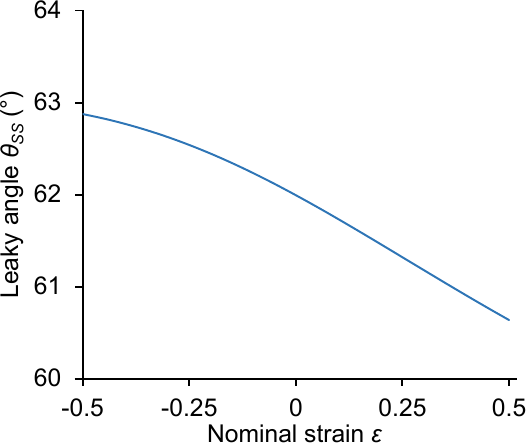}
    \caption{Effect of the prestress on the leaky angle.}
    \label{FigS3}
\end{figure}

The supershear shear surface wave is leaky because of the greater wave speed than the bulk shear wave. According to the Snell's law, the leaky angle $\theta_{SS}$ (see Fig. 4(d)) can be determined by
\begin{equation}
    \cos(\theta_{SS}) = c_T(\theta_{SS}) / c_{SS}.
\label{eq:leaky angle}
\end{equation}
Substitution of Eq. (\ref{eq:incremental_parameters_MR}) into Eq.(\ref{eq:bulk shear wave speed}), we can get $c_T$ as a function of $\theta$. Then the leaky angle can be computed by solving Eq. (\ref{eq:leaky angle}). Figure \ref{FigS3} shows the effect of the prestress on the leaky angle. $\theta_{SS}$ varies from $62.88^{\circ}$ to $60.64^{\circ}$ over a broad range of nominal strain ($-0.5 \le \varepsilon \le 0.5$).

\section{Plane strain Young's modulus and shear modulus of orthotropic material}
Consider the plane-strain state ($\varepsilon_{22} = 0$) of an incompressible, stress-free linear orthotropic material, of which the material symmetric axes are aligned with the coordinate system shown in Fig. 1(a). The Hooke's law that links the the stress and strain is
\begin{equation}
    \sigma_{11} = -\Bar{p} + C_{11}\varepsilon_{11} + C_{13}\varepsilon_{33}, \quad
    \sigma_{33} = -\Bar{p} + C_{13}\varepsilon_{11} + C_{33}\varepsilon_{33}, \quad
    \sigma_{13} = C_{55}\varepsilon_{13},
\end{equation}
where $C_{11}$, $C_{33}$, $C_{13}$, and $C_{55}$ are components of the orthotropic stiffness matrix. $\Bar{p}$ is the Lagrange multiplier for incompressibility constraint. According to the third equation we can get the in-plane shear modulus $\mu_{13}:= \sigma_{13}/\varepsilon_{13} = C_{55}$. Because of the incompressibility we have $\varepsilon_{33} = -\varepsilon_{11}$. Subtracting the second equation from the first equation and taking $\sigma_{33} = 0$ we get $\sigma_{11} = (C_{11} + C_{33} - 2C_{13})\varepsilon_{11}$ and thus the plane strain Young's modulus $E^*_1 := \sigma_{11}/\varepsilon_{11} = (C_{11} + C_{33} - 2C_{13})$. 

Here we introduce an anisotropy index $\Bar{A} = E^*_1/\mu_{13} - 4$. Figure S5 shows the effect of $\Bar{A}$ on the leaky angle.

\begin{figure}[hbt!]
    \centering
    \includegraphics[width=.35\textwidth]{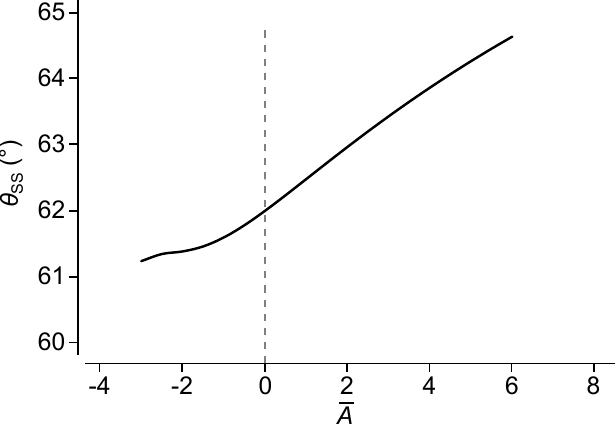}
    \caption{ Variation of the leaky angle $\theta_{SS}$ with the material anisotropy.}
    \label{FigS4}
\end{figure}

\section{Finite element analysis}
The finite element analysis was performed using Abaqus/standard 6.13 (Dassault Systèmes Simulia Corp.). A square domain 40$\times$40 mm$^2$ was built in the analysis, which was large enough to avoid wave reflection. The left side of the domain was a symmetric boundary. A compression/tension along the horizontal direction was applied in a static analysis step. In the subsequent analysis we used an implicit dynamic analysis step and applied a time-harmonic, local pressure to the surface to excite elastic waves. We adopted a gradient mesh ($\sim 0.025$ mm to $\sim 0.5$ mm from top to bottom of the model) and the CPE8RH element type (8-node biquadratic, reduced integration, hybrid with linear pressure).

\clearpage
\section{Experiments}\label{Sec.S3}
\begin{figure}[hbt!]
    \centering
    \includegraphics[width=.75\textwidth]{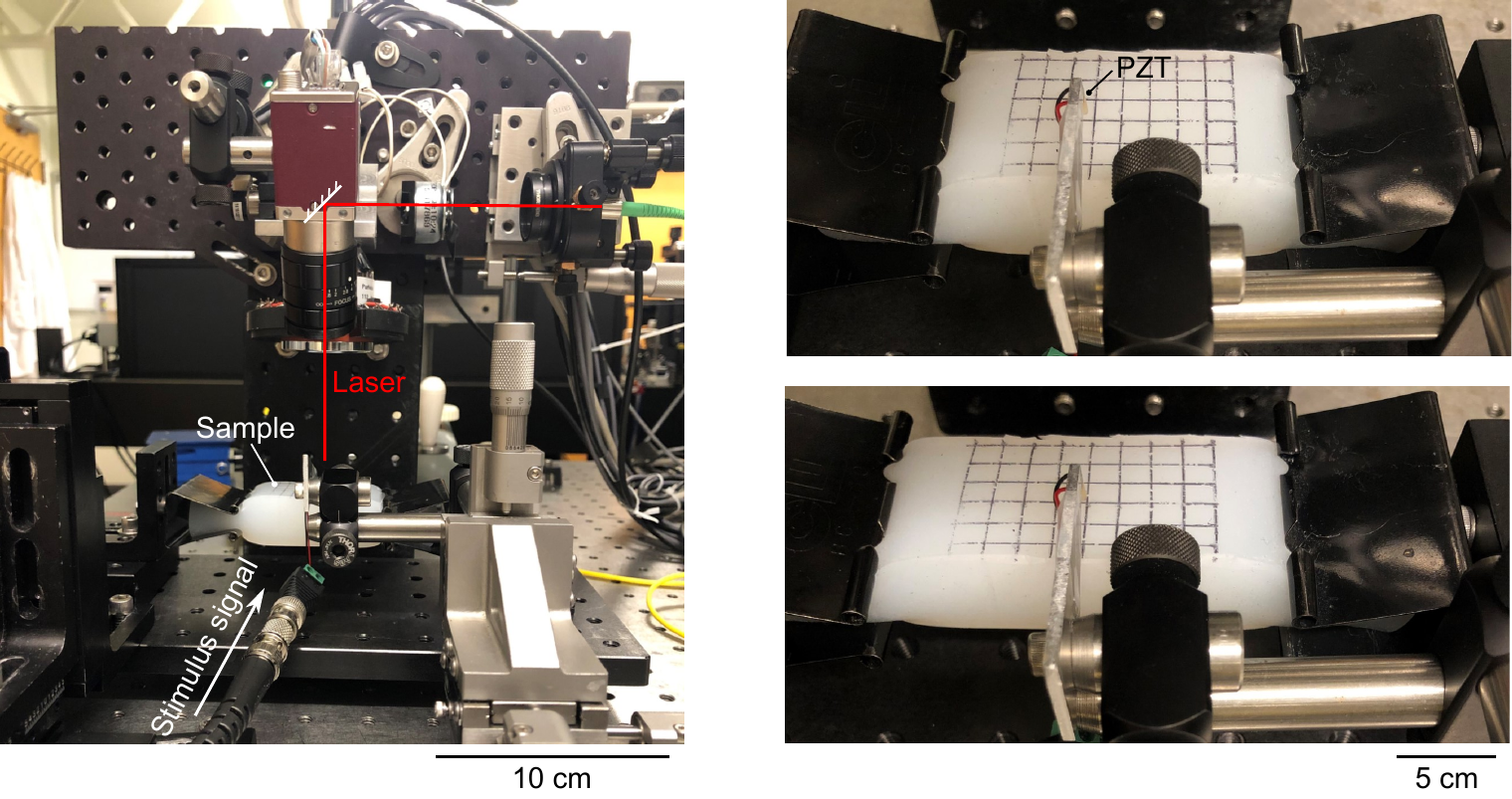}
    \caption{Pictures of the experimental setup.}
    \label{FigS5}
\end{figure}

Figure \ref{FigS5} shows the experimental setup. The sample was clamped and then compressed/stretched manually using a linear translation stage. We drew a mesh grid on the surface of the sample and measured the stretch ratio by the deformation of the mesh grid. The sample was prepared by mixing the two parts with at a 50:50 mass ratio (Ecoflex 0030, Smooth-On Inc.), pouring the mixture into a mold, and then curing the sample at room temperature overnight.

\begin{figure}[hbt!]
    \centering
    \includegraphics[width=.55\textwidth]{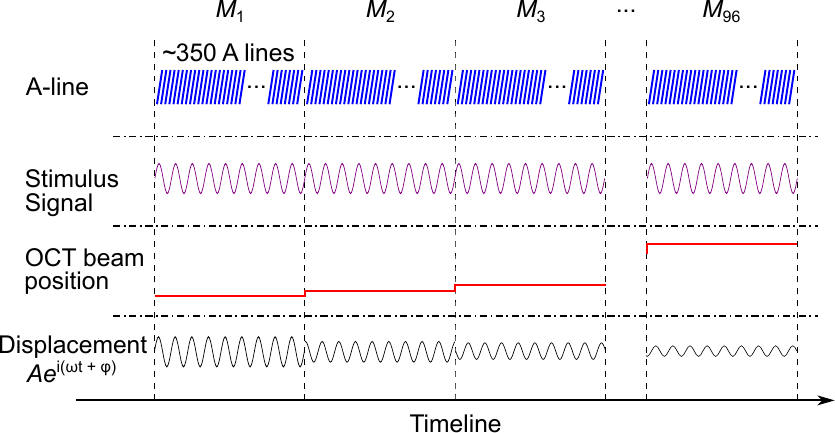}
    \caption{Schematic of the M-B scan.}
    \label{FigS6}
\end{figure}

Figure \ref{FigS6} shows the schematic of the M-B scan. The laser beam scanned synchronously with the stimulus signal sent to the PZT. At each lateral location, we acquired $\sim 350$ A lines (M scan) at a sampling rate of $\sim 43$ kHz. Then the laser beam moved to the next localization (B scan). From each M scan, the vibration on the surface was Fourier transformed to obtain the amplitude $A$ and phase $\varphi$, i.e., $u_z(t) = Ae^{i(\omega t + \varphi)}$. Then we reported the real and imagery parts of the surface displacement.

%